\documentclass[final]{aipproc}

\layoutstyle{6x9}
\pdfoutput=1

\begin{document}

\title{Multi-wavelength Studies of the Gamma-ray Pulsar PSR J1907+0602}

\classification{95.85.Bh; 95.85.Nv; 97.60.Gb}
\keywords{Pulsar, Pulsar Wind Nebula, X-rays, Radio, PSR J1907+0602, MGRO J1908+06, HESS J1908+063}

\author{Dirk Pandel}{
address={Department of Physics, Grand Valley State University, Allendale, MI 49401, USA}
}

\author{Robert Scott}{
address={Department of Physics, Grand Valley State University, Allendale, MI 49401, USA}
}


\begin{abstract}
PSR J1907+0602 is a radio-faint, 107-ms GeV gamma-ray pulsar that was discovered
with the {\it Fermi} LAT in a blind pulsar search.
PSR J1907+0602 is located near the bright, extended TeV gamma-ray source MGRO J1908+06
which may be an associated pulsar wind nebula.
We present an analysis of {\it XMM-Newton} X-ray data and EVLA radio data of the pulsar.
We detect a faint X-ray source coincident with the gamma-ray pulsar and investigate
its spectral and timing properties.
We also find marginal evidence for a bow shock in the X-ray images.
The pulsar was not detected with the EVLA, and we derive upper limits
on the time-averaged radio flux in multiple frequency bands.
\end{abstract}

\maketitle


\section{Introduction}

MGRO J1908+06 (HESS J1908+063) is one of the brightest TeV gamma-ray sources
in the Galactic disk with a flux of $\sim$80\% that of the Crab nebula at 20 TeV.
It was discovered with Milagro \cite{2007ApJ...664L..91A}
and later confirmed with H.E.S.S.\ \cite{2008AIPC.1085..273D}
and VERITAS \cite{2008AIPC.1085..301W} as an extended TeV source
with a radius of $\sim$0.3$^\circ$ \cite{2009A&A...499..723A}.
Even though MGRO J1908+06 is very bright in TeV gamma rays,
no extended emission from the source has so far been detected
at other wavelengths.

In a blind pulsar search, the {\it Fermi} LAT collaboration later discovered
a 107-ms GeV gamma-ray pulsar, PSR J1907+0602, at a position $\sim$0.23$^\circ$
offset from the center of MGRO J1908+06 \cite{2009Sci...325..840A}.
The close proximity of the TeV source to the pulsar suggests that MGRO J1908+06
may be an extended pulsar wind nebula of PSR J1907+0602.
Using a time differencing technique, the position of the gamma-ray pulsar was
determined with arcsecond accuracy, which allowed the detection
of a faint X-ray counterpart with {\it Chandra} as well as faint, pulsed radio emission
with the Arecibo 305-m telescope \cite{2010ApJ...711...64A}.
PSR J1907+0602 has a characteristic are of 19.5~kyr and a spin-down power of
$2.8\times10^{36}\rm\ erg\ s^{-1}$.
We present an analysis of X-ray data from a deep {\it XMM-Newton} observation
of PSR J1907+0602 and of radio data obtained with the Expanded Very Large Array (EVLA).


\section{X-ray Observations}

PSR J1907+0602 was observed with {\it XMM-Newton} \cite{2001A&A...365L...1J}
on 2010 April 26 (Obs.\ ID 0605700201) in one continuous 52 ks observation.
We analyzed the data from the EPIC MOS \cite{2001A&A...365L..27T}
and EPIC PN \cite{2001A&A...365L..18S} cameras
with the {\it XMM-Newton Science Analysis Software} (SAS)
to create X-ray images, spectra, and light curves.

\begin{figure}
\includegraphics[width=2.8in]{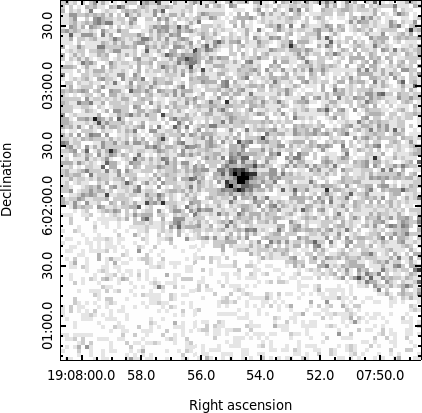}
\hspace{0.1in}
\includegraphics[width=2.8in]{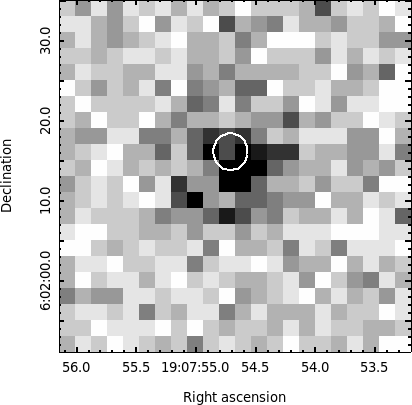}
\caption{
X-ray images of the region around PSR J1907+0602.
Shown are the combined {\it XMM-Newton} data from the three EPIC cameras
in the 1--10 keV energy range.
The images are binned at $2^{\prime\prime}$ and have a size
of $3^\prime$ and $0.75^\prime$ respectively.
The ellipse shows the uncertainty of the pulsar position determined
with the {\it Fermi} LAT \cite{2010ApJ...711...64A}.
Note that the images have not been background subtracted or exposure corrected,
causing the CCD boundary of the EPIC PN to be visible in the left image.
}
\label{xrayimage}
\end{figure}

Figure~\ref{xrayimage} shows X-ray images of the region around PSR J1907+0602.
Shown are the combined {\it XMM-Newton} data from the three EPIC cameras
in the 1--10 keV energy range.
A faint X-ray source is detected at $\rm R.A.=19^h07^m54^s.7$
and $\rm Decl.= 06^\circ02^\prime14^{\prime\prime}.9$
with a statistical uncertainty of $0.8^{\prime\prime}$.
This is consistent with the position of the gamma-ray pulsar determined
with the {\it Fermi} LAT and the position of the X-ray counterpart
detected with {\it Chandra} \cite{2010ApJ...711...64A}.
The apparent size of the X-ray counterpart in the image is consistent
with a point-like source when the angular resolution of {\it XMM-Newton}
is taken into account.
However, the image appears to show a marginal excess $7^{\prime\prime}$
to the lower left of the pulsar which may indicate an extended morphology.
This excess could be interpreted as a bow shock in front of the pulsar
as it is moving away from MGRO J1908+06 (located $\sim$0.2$^\circ$ to the north).
As noted by \cite{2010ApJ...711...64A}, a bow shock would be expected
if the pulsar was born at the center of MGRO J1908+06 and has moved to its current
location during its 19.5~kyr characteristic lifetime.

\begin{figure}
\includegraphics[width=2.84in]{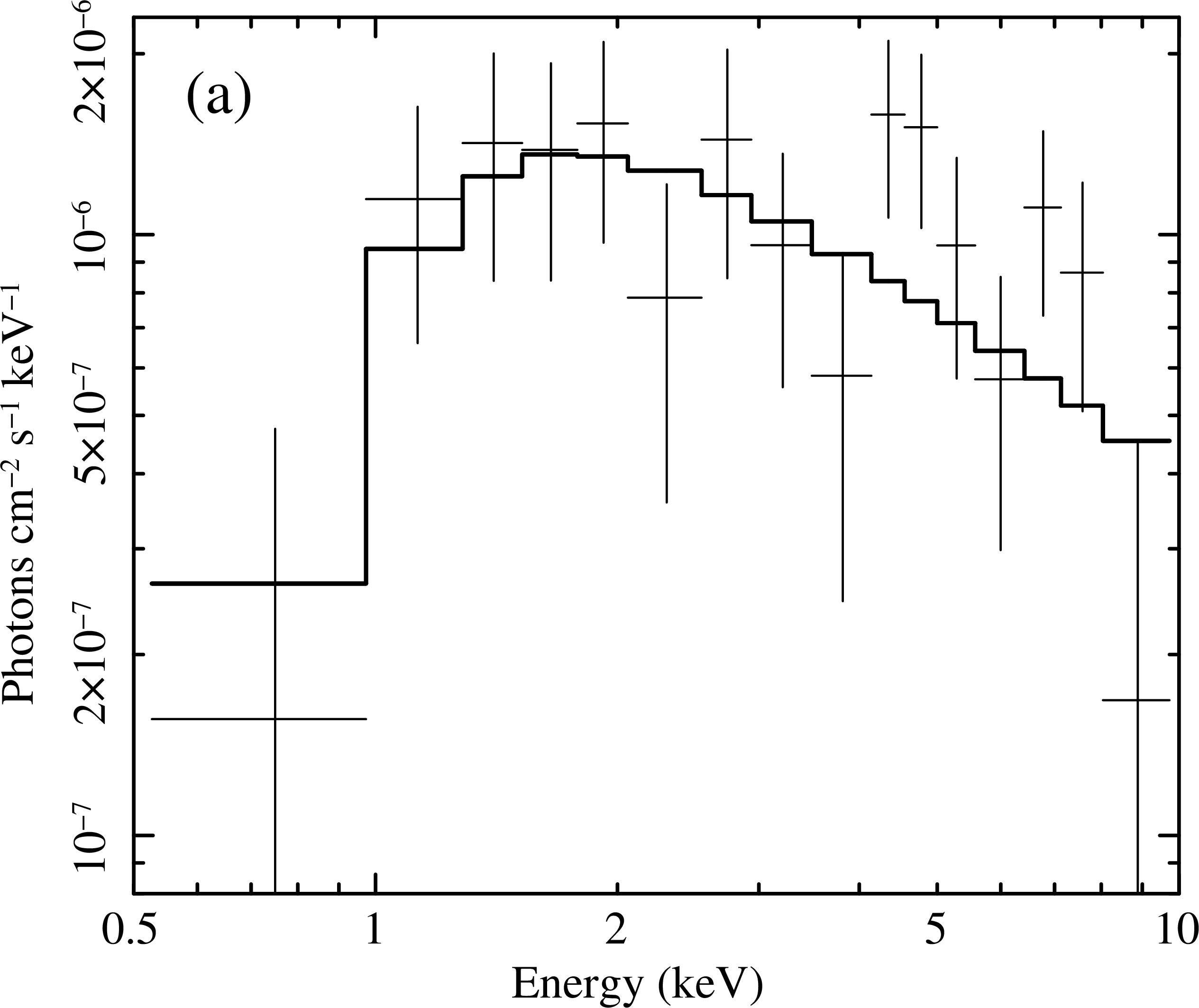}
\hspace{0.1in}
\includegraphics[width=2.76in]{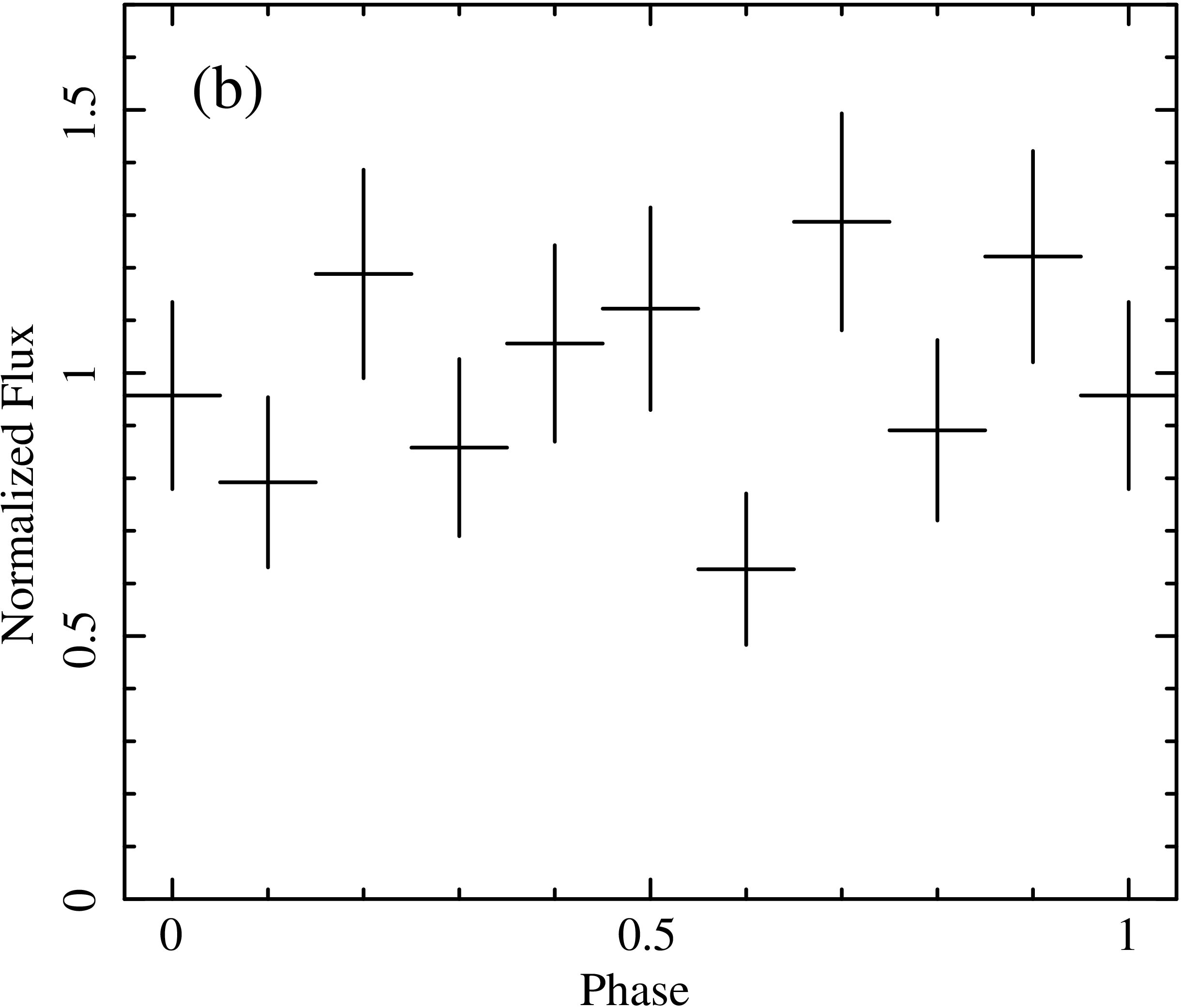}
\caption{
\textbf{(a)} Background-subtracted X-ray spectrum of PSR J1907+0602 from the EPIC PN data.
The solid line shows the best fit with an absorbed power law model.
\textbf{(b)} X-ray light curve of PSR J1907+0602 in the 1--10 keV energy range
folded on the 106.6-ms pulsar spin period.
The light curve has not been background subtracted.
}
\label{xrayspectrum}
\end{figure}

To create X-ray spectra of the pulsar, we extracted events from a circular region
with a radius of $12^{\prime\prime}$.
The spectrum for the EPIC PN is shown in Figure~\ref{xrayspectrum}a.
Significant flux is visible at high energies,
suggesting a non-thermal emission mechanism.
We fitted the spectra from the three EPIC cameras simultaneously
with an absorbed power-law model.
The best-fit parameters are shown in Table~\ref{xrayfit}.
The photon index of $0.9\pm0.4$ is somewhat lower than the typical values
for young pulsars or pulsar wind nebulae.
However, given the characteristic age (19.5 kyr) and spin-down power of the pulsar,
the photon index is consistent with the correlation suggested by
\cite{2003ApJ...591..361G}.
The neutral hydrogen column density obtained from the fit indicates a distance
of several kiloparsec, which is consistent with the 3.2 kpc distance estimate
from radio DM measurements \cite{2010ApJ...711...64A}.
The spectral fit does not require a thermal component from the neutron star surface.
However, such a component is likely hidden by the strong absorption below 1.5~keV.

The pulsar was observed with the EPIC PN in small window mode
which has a sufficiently small timing resolution (6~ms)
for detecting pulsations at the 106.6-ms pulsar spin period.
We searched the X-ray data in the 1--10 keV energy range
for periodic oscillations but did not find significant pulsations
at the spin frequency or any other frequency.
A phase-folded X-ray light curve of PSR J1907+0602 is shown in Figure~\ref{xrayfit}b.
The spectral and timing properties of the source indicate that most of the X-ray emission
originates from a compact pulsar wind nebula that is not resolved in the images.

The close proximity of PSR J1907+0602 to MGRO J1908+06 suggests a likely
association between the two objects.
The 0.23$^\circ$ offset of the pulsar from the center of the TeV source
may be due to a large transverse velocity of the pulsar,
and MGRO J1908+06 may be a relic pulsar wind nebula from
the early life of the pulsar.
Future observations may help to confirm this hypothesis
if a bow shock can be clearly identified and
used to determine the birthplace of the pulsar.

\begin{table}
\begin{tabular}{lcc}
\hline
Neutral Hydrogen Column Density $N_{\rm H}$ ($\rm cm^{-2}$) & & $(0.9\pm0.6)\times10^{22}$ \\
Photon Index & & $0.9\pm0.4$ \\
Absorbed Flux (0.5--10 keV) ($\rm erg\ cm^{-2}\ s^{-1}$) & & $(5.1\pm1.1)\times10^{-14}$ \\
Unabsorbed Flux (0.5--10 keV) ($\rm erg\ cm^{-2}\ s^{-1}$) & & $(5.7\pm1.0)\times10^{-14}$ \\
\hline
\end{tabular}
\caption{Spectral parameters obtained from fitting an absorbed power law model
to the {\it XMM-Newton} data of PSR J1907+0602.
Uncertainties are given at a 90\% confidence level.
}
\label{xrayfit}
\end{table}


\section{Radio Observations}

We have analyzed radio data of PSR J1907+0602 obtained with the EVLA
\cite{2011ApJ...739L...1P} in the frequency bands C, X, and K.
The observations were carried out on 2010 April 30
when the array was in the D configuration.
We analyzed the EVLA data using the {\it Common Astronomy Software Applications}
(CASA) package to create radio maps in the three frequency bands.
We detect no radio source at the location of the pulsar or in its vicinity.
Upper limits on the time-averaged radio flux in the three bands
derived from the RMS noise level are shown in Table~\ref{radiolimits}.
We note that narrow radio pulses have previously been detected at 1.4~GHz
with the Arecibo 305-m telescope \cite{2010ApJ...711...64A}.
The limited timing resolution of the EVLA does not allow the detection
of pulsations at the spin frequency.
The non-detection of a radio counterpart with the EVLA places strong constraints
on the radio emission from any compact pulsar wind nebula surrounding the pulsar.

\begin{table}
\begin{tabular}{ccccc}
\hline
\tablehead{1}{c}{b}{C Band (4.0--8.0 GHz)} & &
\tablehead{1}{c}{b}{X Band (8.0--12.0 GHz)} & &
\tablehead{1}{c}{b}{K Band (18.0--26.5 GHz)} \\
\hline
$5.9\ \rm \mu Jy$ & & $38\ \rm \mu Jy$ & & $36\ \rm \mu Jy$ \\
\hline
\end{tabular}
\caption{Upper limits on the time-averaged radio flux
of PSR J1907+0602 obtained from the EVLA observations.
Limits are given at a 84\% confidence level.}
\label{radiolimits}
\end{table}


\begin{theacknowledgments}
This material is based upon work supported by the U.S.\ National Science Foundation
under Grant No.\ 1068152.
Based on observations obtained with {\it XMM-Newton}, an ESA science mission
with instruments and contributions directly funded by ESA Member States and NASA.
The National Radio Astronomy Observatory is a facility of the National Science Foundation
operated under cooperative agreement by Associated Universities, Inc.
\end{theacknowledgments}


\bibliographystyle{aipproc}
\bibliography{psrj1907}

\end{document}